\documentclass[aps,prl,twocolumn,showpacs,superscriptaddress,floatfix]{revtex4-1}

\usepackage{graphicx}
\usepackage{dcolumn}
\usepackage{bm}
\usepackage{amssymb}
\usepackage{microtype}
\usepackage{xfrac}

\begin{document}

\title{Relief of Frustration in the Heisenberg Pyrochlore Antiferromagnet Gd$_2$Pt$_2$O$_7$}

\author{A.~M.~Hallas}
\affiliation{Department of Physics and Astronomy, McMaster University, Hamilton, ON, L8S 4M1, Canada}

\author{A.~Z.~Sharma}
\affiliation{Department of Chemistry, University of Manitoba, Winnipeg, Manitoba, R3T 2N2, Canada}

\author{Y.~Cai}
\affiliation{Department of Physics and Astronomy, McMaster University, Hamilton, ON, L8S 4M1, Canada}

\author{T.~J.~Munsie}
\affiliation{Department of Physics and Astronomy, McMaster University, Hamilton, ON, L8S 4M1, Canada}

\author{M.~N.~Wilson}
\affiliation{Department of Physics and Astronomy, McMaster University, Hamilton, ON, L8S 4M1, Canada}

\author{M.~Tachibana}
\affiliation{National Institute for Materials Science, 1-1 Namiki, Tsukuba 305-0044, Ibaraki, Japan}

\author{C.~R.~Wiebe}
\affiliation{Department of Physics and Astronomy, McMaster University, Hamilton, ON, L8S 4M1, Canada}
\affiliation{Department of Chemistry, University of Manitoba, Winnipeg, Manitoba, R3T 2N2, Canada}
\affiliation{Department of Chemistry, University of Winnipeg, Winnipeg, MB, R3B 2E9 Canada}
\affiliation{Canadian Institute for Advanced Research, 180 Dundas St. W., Toronto, ON, M5G 1Z7, Canada}

\author{G.~M.~Luke}
\affiliation{Department of Physics and Astronomy, McMaster University, Hamilton, ON, L8S 4M1, Canada}
\affiliation{Canadian Institute for Advanced Research, 180 Dundas St. W., Toronto, ON, M5G 1Z7, Canada}

\date{\today}

\begin{abstract} 
The gadolinium pyrochlores, Gd$_2B_2$O$_7$, are amongst the best realizations of antiferromagnetically coupled Heisenberg spins on a pyrochlore lattice. We present a magnetic characterization of Gd$_2$Pt$_2$O$_7$, a unique member of this family. Magnetic susceptibility, heat capacity, and muon spin relaxation measurements show that Gd$_2$Pt$_2$O$_7$ undergoes an antiferromagnetic ordering transition at $T_N = 1.6$~K. This transition is strongly first order, as indicated by the sharpness of the heat capacity anomaly, thermal hysteresis in the magnetic susceptibility, and a non-divergent relaxation rate in $\mu$SR. The form of the heat capacity below $T_N$ suggests that the ground state is an anisotropic collinear antiferromagnet with an excitation spectrum that is gapped by 0.245(1)~meV. The ordering temperature in Gd$_2$Pt$_2$O$_7$, $T_N = 1.6$~K, is a substantial 160\% increase from other gadolinium pyrochlores, which have been found to order at 1~K or lower. We attribute this enhancement in $T_N$ to the $B$-site cation, platinum, which, despite being non-magnetic, has a filled $5d$ $t_{2g}$ orbital and an empty $5d$ $e_g$ orbital that can facilitate superexchange. Thus, the magnetic frustration in Gd$_2$Pt$_2$O$_7$ is partially ``relieved'', thereby promoting magnetic order.  
\end{abstract}

%\pacs{75.30.Cr, 75.40.Cx, 75.50.Lk}% PACS, the Physics and Astronomy

\maketitle

\section{Introduction}

The pyrochlore oxides, $A_2B_2$O$_7$, are the paragon of geometric magnetic frustration in three dimensions \cite{gardner2010magnetic}. Each of the $A$-site and $B$-site sublattices forms a network of corner sharing tetrahedra. This geometry is highly susceptible to magnetic frustration when either the $A$ or $B$ site is occupied by a magnetic cation. The rare earth pyrochlores are a very interesting subset of these materials, with an astonishing diversity of magnetic ground states and behaviors. This diversity can, in part, be attributed to the different single ion anisotropies realized by the different rare earth cations. Pyrochlores with terbium, dysprosium, or holmium occupying the A-site exhibit Ising spin anisotropy \cite{:/content/aip/journal/jap/87/9/10.1063/1.372565,PhysRevLett.103.056402,2016arXiv160501223R}, while pyrochlores with ytterbium or erbium have XY spin anisotropy \cite{PhysRevB.92.134420,PhysRevLett.103.056402}. Amongst the magnetic rare earth pyrochlores, it is gadolinium alone that provides a good realization of Heisenberg anisotropy on the pyrochlore lattice. This is because gadolinium has a largely isotropic spin only total angular momentum. A number of gadolinium based pyrochlores have now been synthesized and studied, Gd$_2B_2$O$_7$ with $B =$ Ti, Sn, Zr, Hf, and Pb \cite{PhysRevB.59.14489,matsuhira2002low,durand2008heat,PhysRevB.91.104417}. Each of these systems undergoes an antiferromagnetic ordering transition at 1~K or lower.

Heisenberg pyrochlore antiferromagnets have attracted significant interest, beginning with calculations that predicted an infinite ground state degeneracy and, hence, the possibility for spin liquid phenomena \cite{PhysRevLett.80.2933,PhysRevLett.80.2929,PhysRevB.61.1149}. However, it was subsequently shown by Palmer and Chalker that a Heisenberg antiferromagnet with dipolar interactions would order into a four-sublattice state with ordering vector $k = (0,0,0)$ \cite{PhysRevB.62.488}. Investigating these predictions in the gadolinium pyrochlores via neutron scattering is made challenging due to the very high neutron absorption cross-section of gadolinium. In the case of Gd$_2$Ti$_2$O$_7$ \cite{PhysRevB.64.140407} and Gd$_2$Sn$_2$O$_7$ \cite{0953-8984-18-3-L02}, this challenge has been overcome by synthesizing samples with isotopically enriched gadolinium. The ordered state in Gd$_2$Sn$_2$O$_7$ is indeed the $k= (0,0,0)$ ``Palmer-Chalker'' state \cite{0953-8984-18-3-L02}. However, the situation in Gd$_2$Ti$_2$O$_7$ is far more complicated. 

In Gd$_2$Ti$_2$O$_7$, there are two closely separated magnetic ordering transitions, at $T_{N1}=1.0$~K and $T_{N2}=0.75$~K \cite{ramirez2002multiple,bonville2003low,PhysRevB.70.012402}. Despite these two transitions, Gd$_2$Ti$_2$O$_7$ remains in an only partially ordered state down to the lowest measured temperatures. In this partially ordered state, $\sfrac{3}{4}$ of the spins participate in long-range order while the other $\sfrac{1}{4}$ remain in a paramagnetic regime, with short range correlations. Identifying the precise nature of this partially ordered state has proven difficult, as two states, referred to as ``1-$k$'' and ``4-$k$'', are both consistent with most experimental observations \cite{paddison2015nature}. In the 1-$k$ structure, the disordered sites are confined to the triangular layers perpendicular to [111], and the ordered sites reside on the kagome layers with a single propagation vector, $k = (\sfrac{1}{2}, \sfrac{1}{2}, \sfrac{1}{2})$. Conversely, the 4-$k$ structure is a superposition of four $k \in \{\sfrac{1}{2}, \sfrac{1}{2}, \sfrac{1}{2}\}$, with all of the disordered sites confined to share a tetrahedron. The diffuse scattering below $T_{N2}$ has been the primary means of attempting to distinguish between 1-$k$ and 4-$k$. Earlier studies interpreted the diffuse scattering as supporting a picture of 4-$k$ \cite{stewart2004phase}, but more recently this data has been re-analyzed and deemed consistent with only the 1-$k$ structure \cite{paddison2015nature}. The nature of the state intermediate to $T_{N1}$ and $T_{N2}$ is still not definitively known, but a recent theoretical study has shown that thermal fluctuations should select the 4-$k$ structure \cite{javanparast2015fluctuation}. 

In this work, we turn our attention to the magnetism of an overlooked gadolinium pyrochlore, Gd$_2$Pt$_2$O$_7$. While the synthesis of the platinum pyrochlores was first reported nearly 50 years ago \cite{Platinum2,hoekstra1968synthesis,ostorero1974single}, their magnetic characterization has only recently begun \cite{cai2016high}. We investigate the low temperature magnetism of Gd$_2$Pt$_2$O$_7$ using magnetometry, heat capacity, and muon spin relaxation techniques. These probes reveal a transition to an antiferromagnetically ordered ground state in Gd$_2$Pt$_2$O$_7$. Remarkably, this transition occurs at $T_N = 1.6$~K, a substantial enhancement from the other gadolinium-based pyrochlores, which are all found to order at 1~K or lower. We discuss the role of non-magnetic platinum in the relief of frustration in this Heisenberg pyrochlore antiferromagnet.

\section{Synthesis and Experimental Details}

The stability of the pyrochlore lattice, A$_2$B$_2$O$_7$, can be predicted by the ratio of the ionic radii of the $A$ and $B$ site cations. The pyrochlore structure is often stable for $1.4 < R_A/R_B < 2.0$ \cite{wiebe2015frustration}. Given that the ionic radii of Gd$^{3+}$ and Pt$^{4+}$ give a ratio of 1.7, a Gd$_2$Pt$_2$O$_7$ pyrochlore would appear obtainable. Hindering the formation of the pyrochlore phase, however, is the low decomposition temperature of PtO$_2$, only 450$^{\circ}$C, which makes reaction by conventional solid state synthesis impossible. High pressure can be used to suppress the decomposition of PtO$_2$ to sufficiently high temperature, allowing reaction into the pyrochlore phase. Gd$_2$Pt$_2$O$_7$ pyrochlore was prepared from stoichiometric quantities of Gd$_2$O$_3$ and PtO$_2$ using a belt-type high pressure apparatus at 6~GPa and 1000$^{\circ}$C. Small amounts of platinum metal and unreacted Gd$_2$O$_3$ were removed from the reacted product using a solution of boiling aqua regia. The $Fd\bar{3}m$ pyrochlore structure was verified using powder x-ray diffraction with a copper K$_{\alpha 1}$ target, giving a monochromatic beam of x-rays with wavelength $\lambda = 1.5406$~\AA. Rietveld refinement of the measured x-ray diffraction pattern was performed using FullProf \cite{rodriguez1993recent}. 

The bulk magnetic properties of Gd$_2$Pt$_2$O$_7$ were studied with magnetometry and heat capacity measurements. The dc susceptibility was measured using a Quantum Design SQUID magnetometer equipped with a $^3$He insert, allowing measurements to be performed between 0.5~K and 300~K. The heat capacity measurements were collected on warming in a Quantum Design Physical Properties Measurement System with a $^3$He insert, giving a base temperature of 0.4~K. 

Muon spin relaxation ($\mu$SR) measurements on Gd$_2$Pt$_2$O$_7$ were performed at the TRIUMF laboratory in Vancouver, Canada. A 250~mg sample of Gd$_2$Pt$_2$O$_7$ was combined with 100~mg of silver powder, to improve thermal equilibration and mechanical stability, and then pressed into a $\sfrac{3}{8}$" pellet, which was affixed to a silver cold finger using Apiezon N-grease. The $\mu$SR measurements were taken at the M15 surface muon beam line using the Pandora spectrometer and dilution refrigerator, which gives a base temperature of 25~mK and 0.39~ns time resolution. In a $\mu$SR experiment \cite{Tomo}, 100\% spin polarized muons are implanted in the sample, one at a time, and come to rest at Coulomb potential minima. These muons precess in the local magnetic environment and then decay, after an average lifetime of 2.2 $\mu$s, emitting a positron preferentially in the muon spin direction. Histograms of the muon decay events are measured in forwards, $F(t)$, and backwards, $B(t)$ detectors. The muon decay asymmetry is then given by $A(t) = (F(t)-B(t))/(F(t)+B(t))$, which is proportional to the muon spin polarization function. The resulting asymmetry spectra were fitted using the muSRfit software package \cite{suter2012musrfit}.

\section{Powder X-ray Diffraction}

\begin{figure}[tbp]
\linespread{1}
\par
\includegraphics[width=3.3in]{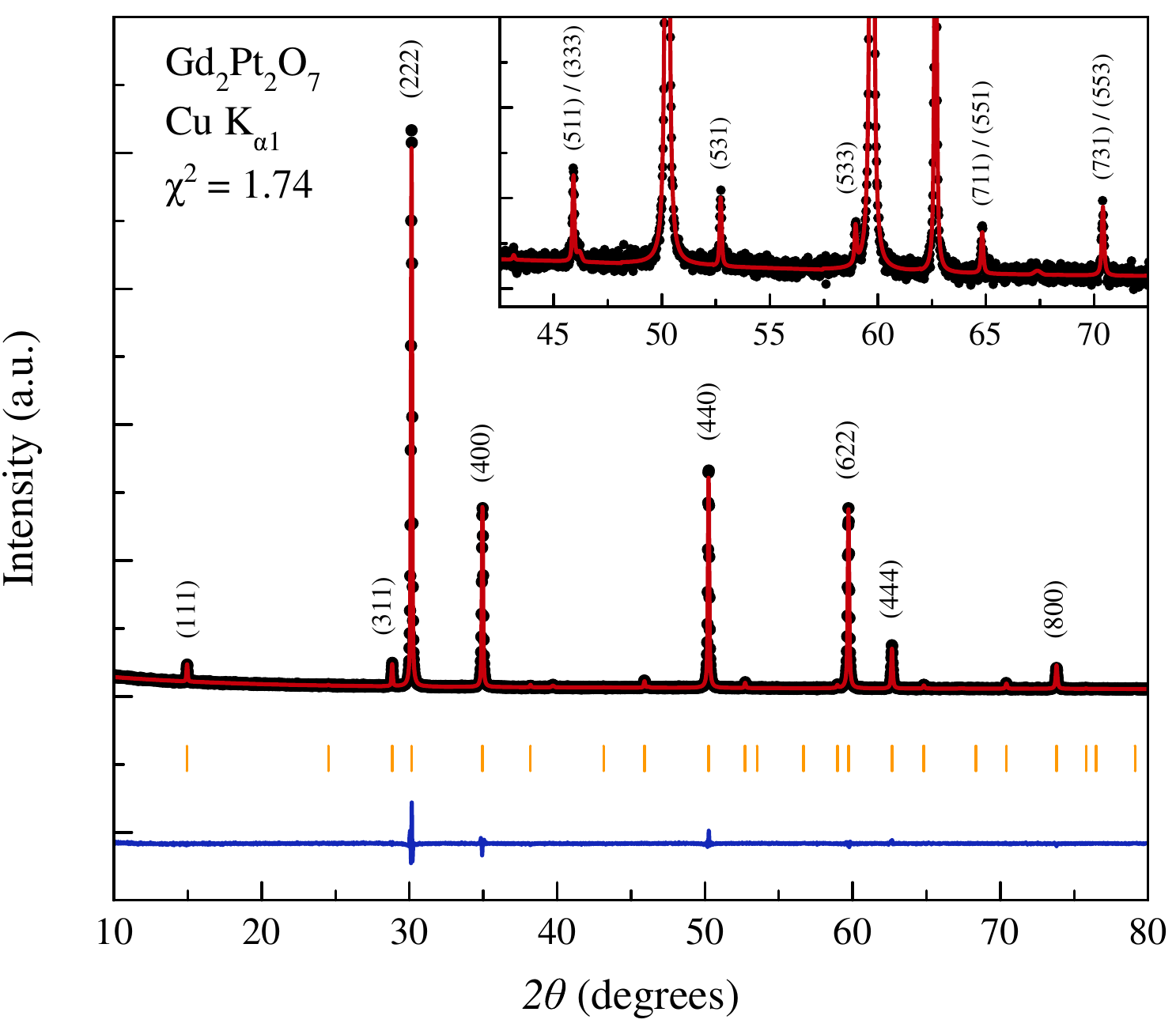}
\par
\caption{The x-ray powder diffraction pattern and Rietveld refinement for Gd$_2$Pt$_2$O$_7$ measured at $T=300$~K with a copper K$_{\alpha 1}$ wavelength, $\lambda = 1.5406$~\AA. The measured data are given by the black points, the fitted curve by the red line and the difference by the blue curve. The allowed nuclear Bragg peak positions are indicated by the yellow marks. The goodness of fit for the refinement into the $Fd\bar{3}m$ space group is $\chi^2 = 1.74$. The inset is an enhanced view of the low intensity peaks between 42.5$^{\circ}$ and 72.5$^{\circ}$ in $2\theta$, showing the absence of any impurity peaks.}
\label{XRD}
\end{figure}

The x-ray powder diffraction pattern and Rietveld refinement for Gd$_2$Pt$_2$O$_7$ is shown in Figure~\ref{XRD}. All of the observed Bragg reflections can be indexed into the $Fd\bar{3}m$ space group. The fitted profile describes the data well, giving a goodness of fit, $\chi^2$, of 1.74. The results of this fit are summarized in Table~\ref{Refinement}. The only adjustable atomic coordinate within the pyrochlore structure is the oxygen $x$ position, which refined to a value of 0.354(1). The lattice constant for Gd$_2$Pt$_2$O$_7$ is $a=10.2626(2)$~\AA, which agrees well with previous reports \cite{Platinum2,hoekstra1968synthesis,ostorero1974single}. 

\begin{table}[tbp]
\begin{tabular}{|l||c|>{\centering}p{1.2cm}|p{1.2cm}<{\centering}|p{1.2cm}<{\centering}|c|}
\toprule
Atom & Wyckoff & $x$ & $y$ & $z$ & B$_{\text{iso}}$ (\AA$^2$) \\
\colrule
Gd & 16$d$ & 0.5 & 0.5 & 0.5 & 0.53(5) \\
%\hline
Pt & 16$c$ & 0 & 0 & 0 & 0.12(4) \\
%\hline
O & 48$f$ & 0.354(1) & 0.125 & 0.125 & 1.8(9) \\
%\hline
O$^{\prime}$ & 8$b$ & 0.375 & 0.375 & 0.375 & 2.1(3) \\
\botrule
\end{tabular}
\caption{Structural parameters from the Rietveld refinement of the x-ray diffraction pattern for Gd$_2$Pt$_2$O$_7$. The only adjustable atomic coordinate is the first oxygen's $x$ position. The lattice parameter refines to a value of $a=10.2626(2)$~\AA~and the goodness-of-fit, $\chi^2$, is 1.74.} %, where $R_{wp} = 10.7$ and $R_{exp} = 8.18$.} 
\label{Refinement}
\end{table}

It is interesting to note that, due to the similar ionic radii of Ti$^{4+}$ and Pt$^{4+}$, the lattice parameters of Gd$_2$Pt$_2$O$_7$ (10.26 \AA) and Gd$_2$Ti$_2$O$_7$ (10.19 \AA~\cite{knop1969pyrochlores}) are quite similar, differing by only 0.7\%. Thus, one might naively expect the magnetic properties of Gd$_2$Pt$_2$O$_7$ to quite closely resemble those of Gd$_2$Ti$_2$O$_7$. However, there are several distinguishing structural properties of Gd$_2$Pt$_2$O$_7$ worth considering. Firstly, Gd$_2$Pt$_2$O$_7$ has a significantly larger oxygen $x$ coordinate than its titanate analog, corresponding to a less distorted oxygen environment about the gadolinium site. Secondly, in other non-magnetic $B$-site pyrochlores of the form Gd$_2B_2$O$_7$, with $B = $ Ti \cite{knop1969pyrochlores}, Sn \cite{kennedy1997structural}, Zr \cite{kennedy2011neutron}, Hf \cite{durand2008heat}, and Pb \cite{PhysRevB.91.104417}, the $B$-site cation has a closed shell electron configuration. Platinum differs in this regard, with a [Xe]5$d^6$ valence shell, leading to a filled $t_{2g}$ orbital and an empty $e_g$ orbital. We will discuss this difference further, in the context of the magnetic properties of Gd$_2$Pt$_2$O$_7$, below. 

\section{Magnetic Susceptibility}

\begin{figure}[tbp]
\linespread{1}
\par
\includegraphics[width=3.3in]{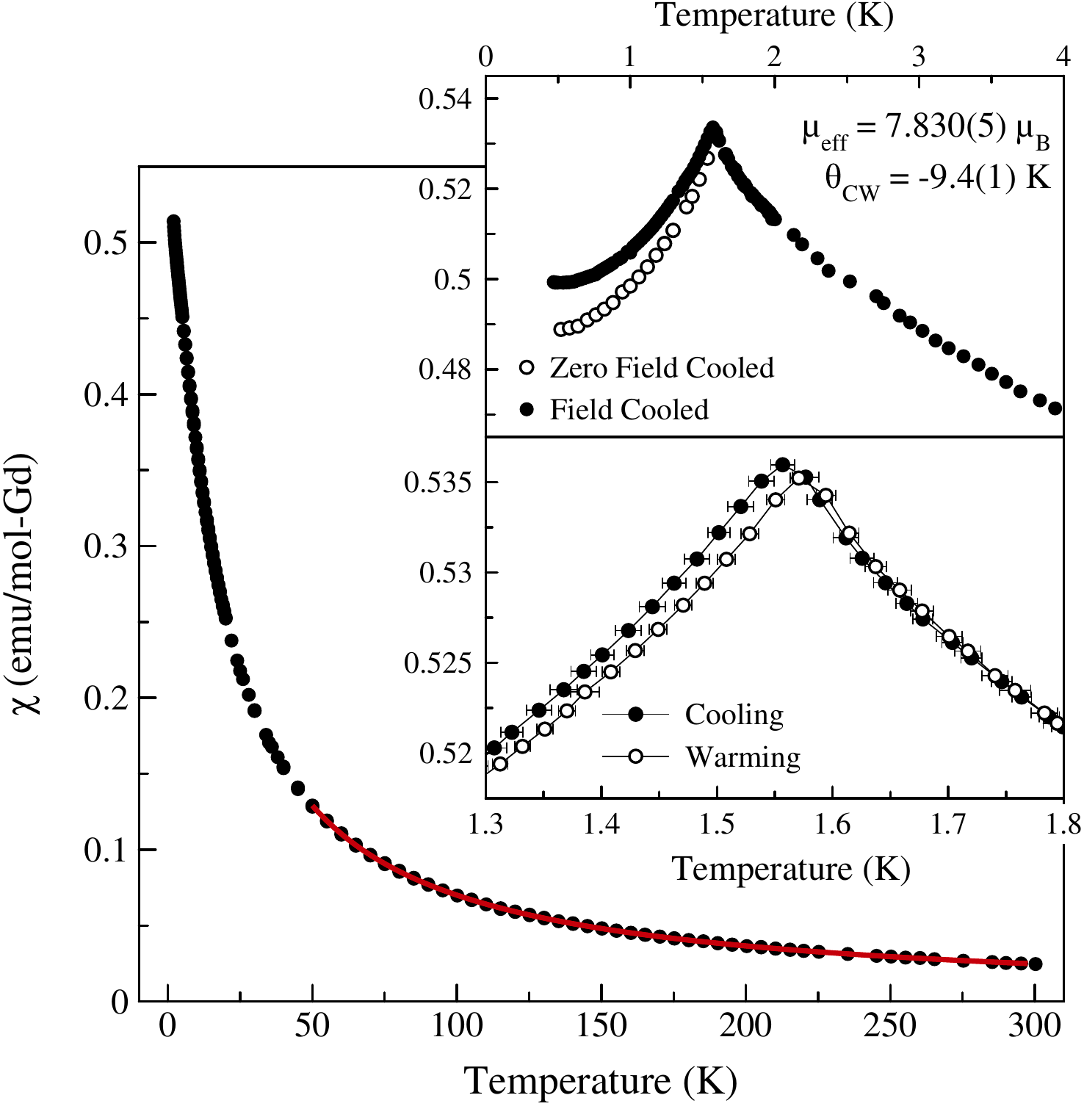}
\par
\caption{The dc magnetic susceptibility of Gd$_2$Pt$_2$O$_7$ measured in an $H=0.1$~T applied field. The high temperature data, between 50~K and 300~K, is well-described by the Curie-Weiss law, which is given by the red curve. Upper Inset: The low temperature dc susceptibility, between 0.5~K and 5~K. An antiferromagnetic ordering transition is observed at 1.6~K, as signaled by the cusp and the bifurcation of the field-cooled and zero-field-cooled susceptibility. Lower Inset: Field cooled measurements taken on warming and cooling show a thermal hysteresis of 30~mK below the maximum of the transition.}
\label{Susceptibility}
\end{figure}

The dc magnetic susceptibility of Gd$_2$Pt$_2$O$_7$ measured in a 0.1~T applied field is shown in Figure~\ref{Susceptibility}. At high temperature, the susceptibility is well-described in terms of the Curie-Weiss law. Fits between 50~K and 300~K, give an antiferromagnetic Curie-Weiss temperature, $\theta_{\text{CW}}$, of $-9.4(1)$~K. The fitted effective paramagnetic moment, $\mu_{\text{eff}}$, is 7.830(5)~$\mu_B$, close to the 7.94~$\mu_B$ free-ion moment. As Gd$^{3+}$ has seven $f$ electrons (\emph{ie.} a half-filled $f$ electron shell), the orbital angular momentum is quenched, $L=0$, resulting in an isotropic spin-only total angular momentum of $S=J=\sfrac{7}{2}$. Thus, crystal electric field effects are expected to be relatively unimportant, as it is the anisotropic orbital angular momentum that is primarily responsible for splitting the $2J+1 = 8$-fold ground state multiplet. Consequently, the Curie-Weiss temperature can be realistically interpreted as a measure of the exchange interactions, as opposed to crystal field effects as in many other rare earth pyrochlores. 

At 1.6~K, Gd$_2$Pt$_2$O$_7$ undergoes an antiferromagnetic ordering transition, marked by a sharp cusp in the susceptibility (Figure~\ref{Susceptibility} Upper Inset). Taken with the Curie-Weiss constant, we can then determine the frustration index for Gd$_2$Pt$_2$O$_7$, $f = \theta_{\text{CW}}/T_N \approx 6$, which corresponds to only moderate frustration \cite{greedan2001geometrically}. Below $T_N = 1.6$~K, there is a divergence of the zero-field-cooled and field-cooled susceptibilities, indicative of glassy dynamics or domain effects. A similar effect is observed in both Gd$_2$Sn$_2$O$_7$, which has an antiferromagnetic Palmer-Chalker ground state \cite{bonville2003low,0953-8984-18-3-L02}, and in Er$_2$Ti$_2$O$_7$, which has an antiferromagnetic $\psi_2$ ground state \cite{0953-8984-12-4-308,PhysRevB.68.020401}. Conversely, there is a negligibly small field-cooled/zero-field-cooled splitting in Gd$_2$Ti$_2$O$_7$, which has a highly unusual partially ordered state \cite{paddison2015nature}. Lastly, within the field cooled protocol, measurements were performed passing through the anomaly at $T_N=1.6$~K on warming and cooling (Figure~\ref{Susceptibility} Lower Inset). Below $T_N=1.6$~K, there is a thermal hysteresis of 30~mK, significantly larger than the instrumental resolution. This is clear evidence that this magnetic ordering transition is first order. 

%--------------------------------------------------------------------------------------------------

\section{Heat Capacity}

\begin{figure}[tbp]
\linespread{1}
\par
\includegraphics[width=3.3in]{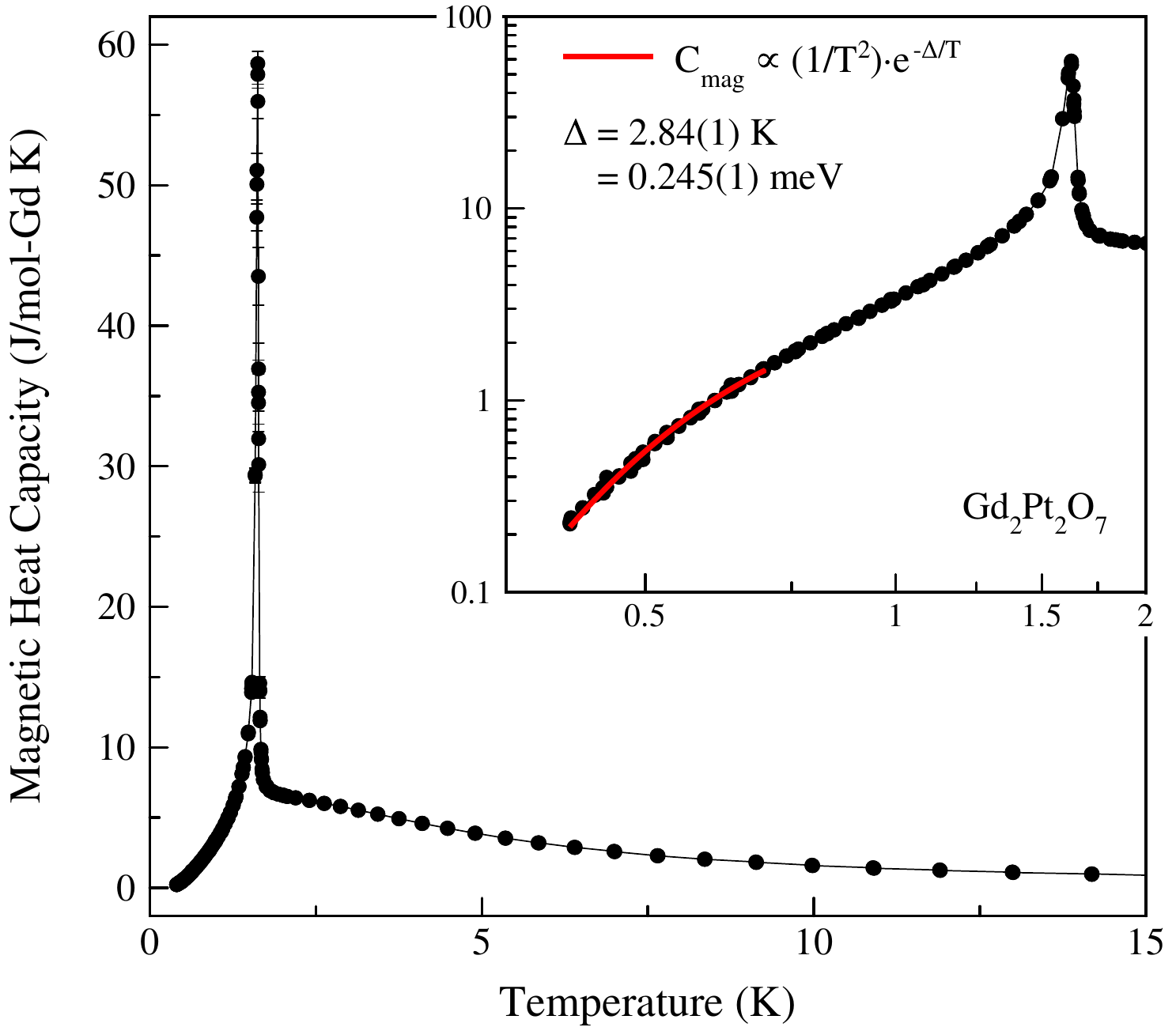}
\par
\caption{The low temperature magnetic heat capacity of Gd$_2$Pt$_2$O$_7$, showing a first order phase transition at $T_N = 1.6$~K. The lattice contribution, while very small below 15~K, was subtracted using the heat capacity of non-magnetic isostructural Lu$_2$Pt$_2$O$_7$. Inset: The low temperature heat capacity, between 0.4~K and 2~K on a logarithmic scale. Below 0.75~K, the heat capacity is well-described by an exponential decay, given by the red curve, indicating the presence of a gap in the spin excitation spectrum. The fitted value of the gap is $\Delta = 0.245(1)$~meV.}
\label{HeatCapacity}
\end{figure}

The magnetic heat capacity of Gd$_2$Pt$_2$O$_7$, shown in Figure~\ref{HeatCapacity}, also shows a magnetic ordering transition at $T_N = 1.6$~K. This sharp transition appears first order, with its maximum at 58.6~J/mol-Gd K. The calculated magnetic entropy, determined by integrating $C_{\text{mag}}/T$ up to 15~K, gives the full $R\ln{8}$ entropy, as expected for $J=\sfrac{7}{2}$. Notably, there is only a single transition in Gd$_2$Pt$_2$O$_7$, unlike Gd$_2$Ti$_2$O$_7$ where there are two distinct ordering transitions \cite{ramirez2002multiple,bonville2003low,PhysRevB.70.012402}. Instead, the specific heat of Gd$_2$Pt$_2$O$_7$ more strongly resembles that of Gd$_2$Sn$_2$O$_7$, which also has a single first order transition \cite{bonville2003low}.

More can be learned about the ordered state in Gd$_2$Pt$_2$O$_7$ by considering the magnetic heat capacity below $T_N$, a probe of the spin excitations arising from the ordered state. Its worth noting that gadolinium does have a nuclear contribution to its heat capacity. However, it only becomes substantial below 200~mK \cite{quilliam2007evidence}, which is outside the measured range in this study and thus, we need not consider it. Below $T_N$, the heat capacity in Gd$_2$Pt$_2$O$_7$ does not follow a $C_{\text{mag}} \propto T^3$ dependence, as would be expected for conventional ungapped antiferromagnetic spin waves. Nor does it follow a $C_{\text{mag}} \propto T^2$ dependence, which is the unusual temperature dependence observed in Gd$_2$Ti$_2$O$_7$ \cite{PhysRevLett.95.047203}. Instead, the lowest temperature magnetic heat capacity of Gd$_2$Pt$_2$O$_7$ can be well parameterized by $C_{\text{mag}} \propto \left(\sfrac{1}{T^2}\right)e^{-\Delta/T}$, which is the expected form for a gapped spin wave spectrum in a collinear antiferromagnet with anisotropy. The fitted value of the gap, $\Delta$, is 0.245(1)~meV. This once again mirrors the situation in Gd$_2$Sn$_2$O$_7$, where the ordering temperature, $T_N = 1.0$~K, and the magnitude of the spin wave gap, $\Delta = 0.121(1)$~meV, is much smaller than in Gd$_2$Pt$_2$O$_7$ \cite{quilliam2007evidence}. 

%--------------------------------------------------------------------------------------------------

\section{Muon Spin Relaxation}

\begin{figure}[tbp]
\linespread{1}
\par
\includegraphics[width=3.3in]{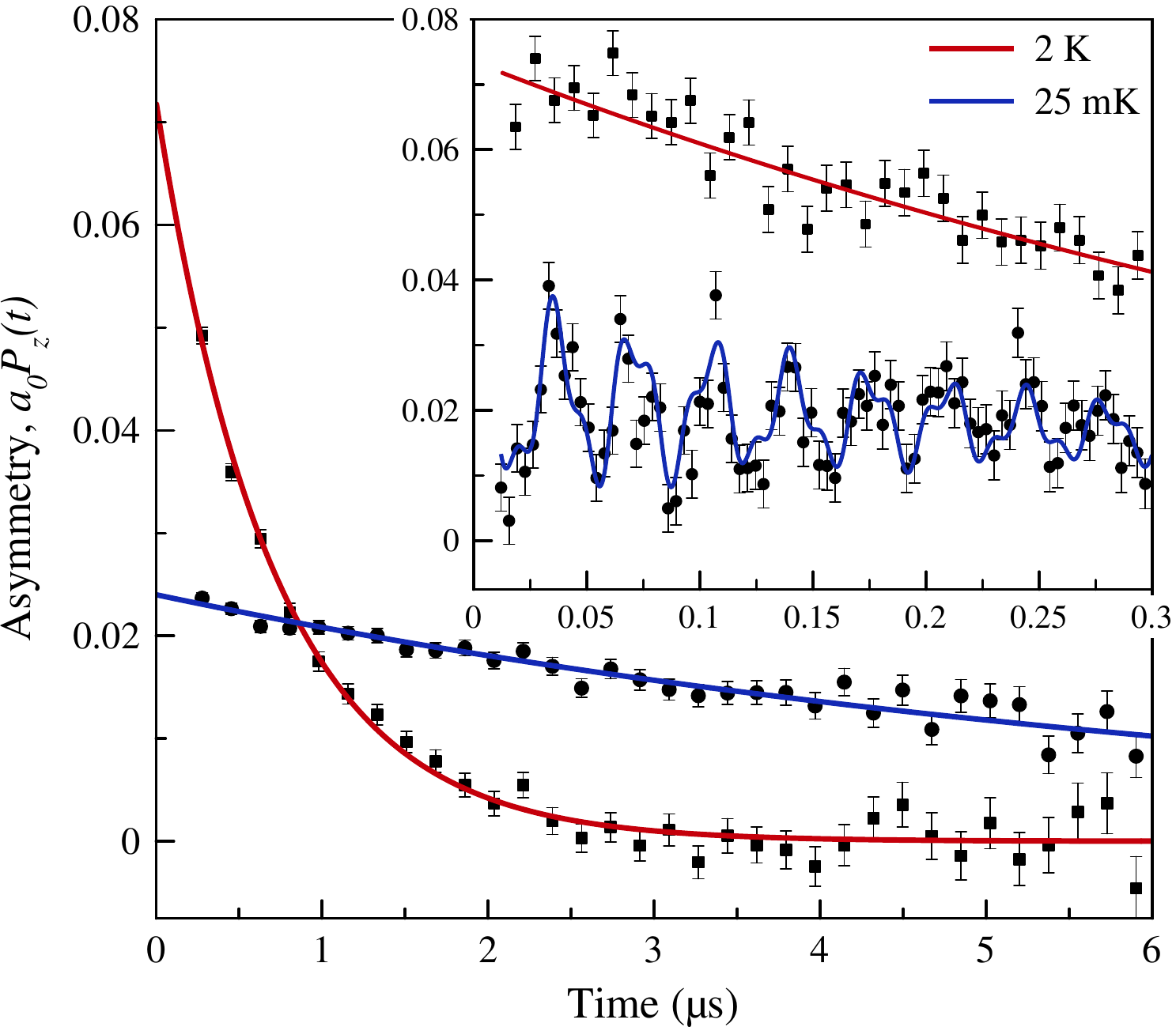}
\par
\caption{The muon decay asymmetry Gd$_2$Pt$_2$O$_7$ out to 6~$\mu$s at 25~mK and 2~K. Below $T_N=1.6$~K, the relaxation rate steeply increases such that the early time asymmetry falls outside of our time window at 25~mK. The fits to the data are given by an exponential decay. Inset: The early time asymmetry, below 0.3~$\mu$s, where oscillations can be resolved at 25~mK. The fit at 25~mK is the sum of two cosine functions with frequencies of 0.21(1)~T and 0.49(2)~T.}
\label{muSR}
\end{figure}

The muon decay asymmetry in Gd$_2$Pt$_2$O$_7$ in a very small 10~G longitudinal field is shown above and below $T_N = 1.6$~K in Figure~\ref{muSR}. The purpose of the 10~G field was to decouple any relaxation due to nuclear dipole moments from the sample or silver sample holder. Below $T_N = 1.6$~K, the initial asymmetry is dramatically reduced. This indicates that very fast relaxation occurs within the first 10~ns, which is the deadtime that follows the initial muon implanatation and is thus outside of our time window. Fast relaxation like this is commonly observed in materials with very large magnetic moments, such as other gadolinium-based systems \cite{PhysRevB.73.172418,Chapuis2009686}. Thus, passing through the magnetic ordering transition, the relaxation rate of the front end rapidly increases.

Examining more closely the early time, the first 0.3~$\mu$s, shown in the inset of Figure~\ref{muSR}, we observe an oscillatory component to the asymmetry below $T_N$. An oscillating muon decay asymmetry is a hallmark of static magnetic order. The muon decay asymmetry in Gd$_2$Pt$_2$O$_7$ at 25~mK fits to an exponential relaxation and two damped cosines:
\begin{eqnarray}
A(t) = A_1e^{-\lambda_1t}\cos(\gamma_{\mu}B_1t+\phi) +
\nonumber \\
A_2e^{-\lambda_2t}\cos(\gamma_{\mu}B_2t+\phi) + A_3e^{-\lambda_3t}
\end{eqnarray}
The pyrochlore structure has two crystallographically unique oxygen sites, which naturally correspond to two distinct muon stopping sites, and hence, two oscillation frequencies. The fits at 25~mK indicate that the internal field strength at these two muon stopping sites are $B_1 = 0.21(1)$~T and $B_2 = 0.49(2)$~T, similar in magnitude to the internal fields found in Gd$_2$Sn$_2$O$_7$ \cite{Chapuis2009686}. The application of a longitudinal magnetic field at 25~mK partially decouples the fast relaxing asymmetry at $H=0.1$~T, and fully decouples it at $H=1$~T, as would be expected for conventional static order.

\begin{figure}[tbp]
\linespread{1}
\par
\includegraphics[width=3.3in]{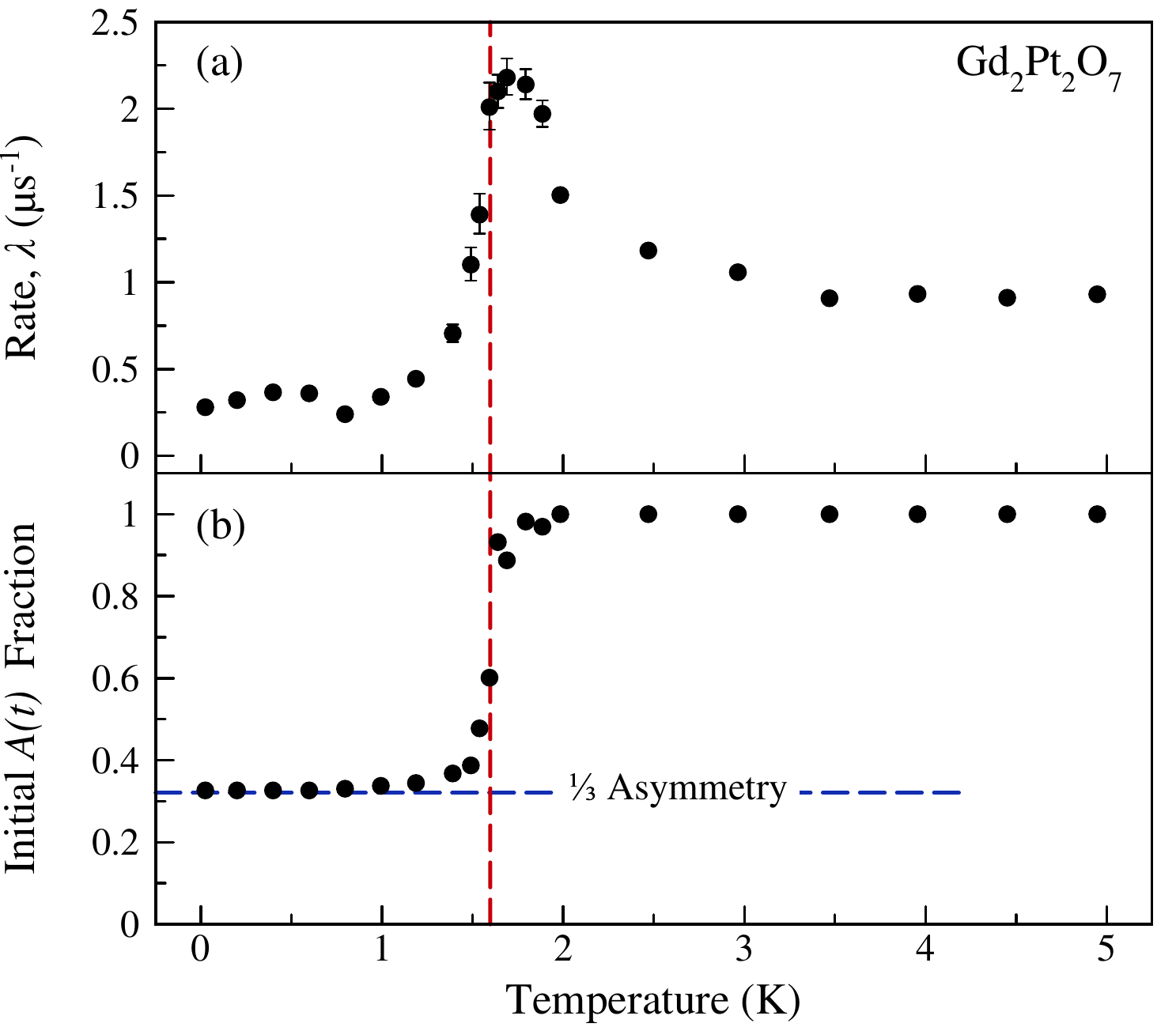}
\par
\caption{Temperature dependence of two fit parameters, (a) the relaxation rate, $\lambda$, and (b) the fraction of the initial asymmetry, for the fits to the muon decay asymmetry in Gd$_2$Pt$_2$O$_7$. The relaxation rate is peaked at $T_N = 1.6$~K, while the initial asymmetry falls off sharply at $T_N=1.6$~K, which is marked by the dashed red line. The initial asymmetry in the magnetically ordered state is $\sfrac{1}{3}$ of the total asymmetry, as marked by the dashed blue line, where the remaining $\sfrac{2}{3}$ asymmetry is relaxing so fast that it falls outside of our time window.} 
\label{muSR_OP}
\end{figure}

The long time asymmetry of Gd$_2$Pt$_2$O$_7$ is well-described by a single exponential relaxation function (Figure~\ref{muSR}). The relaxation rate, $\lambda$, as a function of temperature is peaked at $T_N=1.6$~K (Figure~\ref{muSR_OP}(a)) but does not diverge. A divergent relaxation rate at $T_N$ is expected for a second order transition, and the absence of such a divergence is then further demonstration of the first order nature of this magnetic phase transition \cite{FirstOrderUemura}. Below $T_N$, there is a residual relaxation rate of approximately 0.3~$\mu$s$^{-1}$ that persists to the lowest measured temperature. In a conventional magnetic ordering transition, the relaxation rate well-below $T_N$ should approach zero. Experimental evidence, however, would suggest that a non-zero relaxation rate at the lowest measured temperatures is a characteristic feature of magnetically frustrated materials. This effect is attributed to persistent spin dynamics. In particular, persistent spin dynamics have been observed in other frustrated gadolinium-based systems \cite{PhysRevLett.85.3504,PhysRevB.73.172418}, including Gd$_2$Sn$_2$O$_7$ \cite{Chapuis2009686}, as well as other pyrochlores with long-range magnetic order, such as Er$_2$Ti$_2$O$_7$ \cite{0953-8984-17-6-015} and Tb$_2$Sn$_2$O$_7$ \cite{PhysRevLett.96.127202,PhysRevLett.97.117203}. While an understanding of persistent spin dynamics is far from complete \cite{0953-8984-23-16-164216,PhysRevB.91.104427}, they are, regardless, ubiquitous in geometrically frustrated systems. The persistent spin dynamics in Gd$_2$Pt$_2$O$_7$ are not fully decoupled by even an $H = 1$~T field, which is far larger than the magnitudes of the internal fields.

Finally, we can consider the amount of asymmetry that rapidly relaxes outside of our time window, or ``lost asymmetry'', as a function of temperature. In Figure~\ref{muSR_OP}(b) we plot the initial asymmetry, expressed as a fraction of the full asymmetry. The full asymmetry is defined by the value of the initial asymmetry in the paramagnetic regime at 2~K and above, which is $a_0 = 0.072(1)$. Below $T_N=1.6$~K, the initial asymmetry falls off precipitously. At the lowest temperatures, between 1~K and 25~mK, only $\sfrac{1}{3}$ of the initial asymmetry is detected, where the remaining $\sfrac{2}{3}$ is part of the rapidly relaxing front end, and falls outside our time window. This division of the muon decay asymmetry into $\sfrac{1}{3}$ and $\sfrac{2}{3}$ components is the result of the isotropic environment produced in a powder sample, where directional averaging gives a $\sfrac{1}{3}$ contribution longitudinal to the initial muon spin direction and a $\sfrac{2}{3}$ transverse contribution. The $\sfrac{1}{3}$ component gives the long time relaxation. The $\sfrac{2}{3}$ contribution accounts for both the oscillating component and the rapidly relaxing front end. The front end relaxes within the first 10~ns, indicating a relaxation rate in excess of 100~$\mu$s$^{-1}$.

%--------------------------------------------------------------------------------------------------

\section{Discussion and Conclusions}

\begin{figure}[tbp]
\linespread{1}
\par
\includegraphics[width=2.6in]{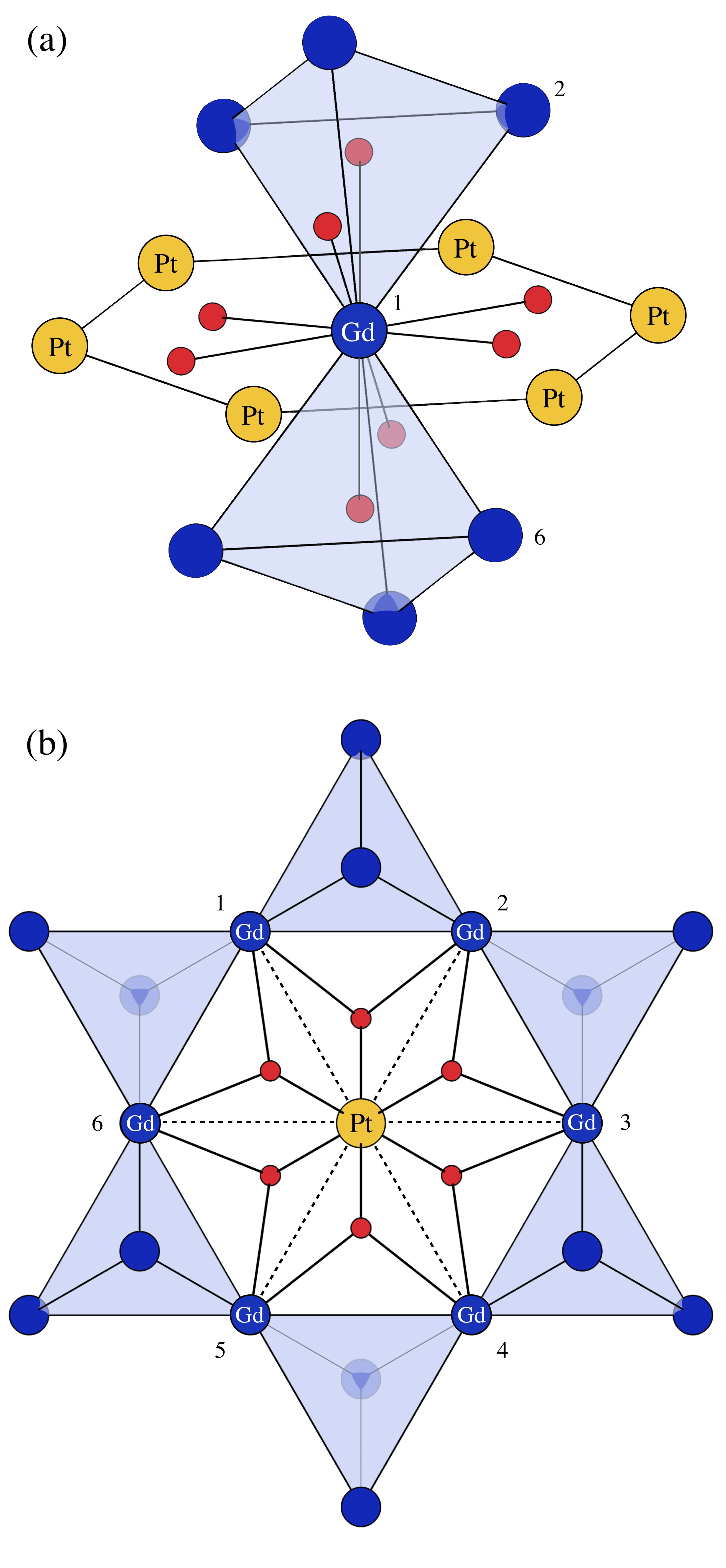}
\par
\caption{Schematic views of the local environments of (a) gadolinium and (b) platinum in Gd$_2$Pt$_2$O$_7$, where the gadolinium cations are given in blue, the platinum cations in yellow and the oxygen anions in red. (a) Gadolinium sits in an eight-fold coordinate oxygen environment with six nearest neighbor platinum cations. (b) Looking along the [111] crystallographic direction, the Gd$-$O$-$Pt$-$O$-$Gd exchange pathways can be clearly observed.}
\label{Structure}
\end{figure}

In many regards, Gd$_2$Pt$_2$O$_7$ strongly resembles its sister compound Gd$_2$Sn$_2$O$_7$: the clear antiferromagnetic transition in its susceptibility, the strongly first order heat capacity transition and gapped spin wave spectrum. In fact, these signatures all suggest that Gd$_2$Pt$_2$O$_7$ is a good candidate for the Palmer-Chalker ground state \cite{PhysRevB.62.488,0953-8984-18-3-L02}. However, this determination will likely have to await a neutron diffraction measurement, which would require a sample prepared with isotopically enriched gadolinium. 

There are, however, several distinctive properties in Gd$_2$Pt$_2$O$_7$ that bear consideration. In particular, the Neel ordering transition, $T_N = 1.6~$K, is substantially increased over the values of 1~K or lower, as seen in other Gd-pyrochlores. As the lattice parameters of Gd$_2B_2$O$_7$ ($B=$ Ti, Pt, Sn) are all quite similar, the difference in the dipolar interaction is likely negligible. Instead, we attribute this enhancement to the non-magnetic platinum which occupies the $B$-site. The empty platinum $5d$ $e_g$ orbitals open additional, and apparently important, superexchange pathways. We propose that this superexchange pathway would likely involve the gadolinium $5d$ orbitals, as the $4f$ orbitals are too spatially localized. The superexchange pathway would then proceed from Gd$-$O$-$Pt$-$O$-$Gd. Of the eight oxygens surrounding each gadolinium, all except for the axial oxygens are shared with nearest neighbor platinum cations. The six nearest neighbor platinums form a hexagon surrounding gadolinium in the plane perpendicular to the local $<$111$>$ axis, as can be seen in Figure~\ref{Structure}(a). Referring to Figure~\ref{Structure}(b), it can be seen that while Gd$1$ shares an oxygen with Gd$2$ and Gd$6$, it does not with Gd$3$, Gd$4$ or Gd$5$. Thus, our proposed superexchange pathway, Gd$-$O$-$Pt$-$O$-$Gd, would then be relevant for each of these three Gd sites, which is repeated for each of the six platinum nearest neighbors. 

While this proposal remains speculative, corroborative evidence comes from the other Pt pyrochlores, Yb$_2$Pt$_2$O$_7$ and Er$_2$Pt$_2$O$_7$, where the ordering temperatures are respectively higher and lower than what would be expected from the lattice parameter trend \cite{cai2016high}. The exact effect of this additional superexchange pathway would be dependent on the anisotropic exchange parameters, which vary a great deal from system to system. In Yb$_2$Pt$_2$O$_7$, the net effect is to enhance ferromagnetic order, while in Er$_2$Pt$_2$O$_7$ the net effect is to suppress antiferromagnetic order \cite{cai2016high}. In the present case of Gd$_2$Pt$_2$O$_7$, the effect is to \emph{enhance} antiferromagnetic order. Thus, the detailed manifestation of this superexchange pathway in the platinum pyrochlores remains an open question, which could possibly be addressed using density functional theory. 

In summary, we have found that replacement of the $B$-site in the gadolinium pyrochlores with platinum results in an effective reduction in the geometric magnetic frustration. Gd$_2$Pt$_2$O$_7$ undergoes a strongly first order antiferromagnetic transition at $T_N=1.6$~K. While the energy scales of rare earth magnetism are small as compared to transition metal systems, it is worth highlighting that this is a substantial 160\% enhancement from other gadolinium pyrochlores. This reduction in frustration can be understood in terms of the superexchange pathway via the empty $5d$ $e_g$ orbitals in non-magnetic platinum. This study therefore offers a fresh insight into the phenomena of geometric magnetic frustration by exploring a novel mechanism through which the frustration can be tuned. 

\begin{acknowledgments}
A.M.H. thanks J.~E.~Greedan for helpful conversations regarding the role of platinum in the magnetic properties of this system and J.~Gaudet for useful discussions. We greatly appreciate the hospitality of the TRIUMF Centre for Molecular and Materials Science and the technical support of B.S. Hitti and D.J. Arseneau throughout the $\mu$SR measurements. A.M.H. acknowledges support from the Vanier Canada Graduate Scholarship Program and thanks the National Institute for Materials Science (NIMS) for their hospitality and support through the NIMS Internship Program. This work was supported by the Natural Sciences and Engineering Research Council of Canada and the Canada Foundation for Innovation. C.R.C. acknowledges support from the Canada Research Chairs program (Tier II).
\end{acknowledgments}

\bibliography{GdPtO_Ref}

\end{document}